\documentclass[conference]{IEEEtran}
\IEEEoverridecommandlockouts
\usepackage{cite}
\usepackage{amsmath,amssymb,amsfonts}
\usepackage{algorithmic}
\usepackage{graphicx}
\usepackage{textcomp}
\usepackage{xcolor}

\usepackage{comment}
\usepackage{multirow}
\usepackage{graphicx}
\usepackage{booktabs} 
\usepackage{fancybox}
\usepackage{fixltx2e}
\usepackage{amsmath}
\usepackage{tabularx}
\usepackage{amsthm}
\usepackage{amssymb}
\usepackage{url}

\usepackage{xcolor}
\usepackage{listings}
\PassOptionsToPackage{bookmarks=false}{hyperref}

\makeatletter

\ifCLASSINFOpdf
\else
\fi

\def\BibTeX{{\rm B\kern-.05em{\sc i\kern-.025em b}\kern-.08em
    T\kern-.1667em\lower.7ex\hbox{E}\kern-.125emX}}

\begin{document}



\title{Governance \& Autonomy:\\ Towards a Governance-based Analysis of Autonomy in Cyber-Physical Systems-of-Systems\thanks{\textcopyright   \, 2020 IEEE Personal use of this material is permitted. Permission from IEEE must be obtained for all other uses, in any current or future media, including reprinting/republishing this material for advertising or promotional purposes, creating new collective works, for resale or redistribution to servers or lists, or reuse of any copyrighted component of this work in other works} }




\author{\IEEEauthorblockN{Mohamad Gharib, Paolo Lollini, Andrea Ceccarelli, Andrea Bondavalli}
\IEEEauthorblockA{University of Florence - DiMaI\\
Viale Morgagni 65, Florence, Italy\\
\{mohamad.gharib,paolo.lollini,andrea.ceccarelli,andrea.bondavalli\}@unifi.it}}

\maketitle


\begin{abstract}

   One of the main challenges in integrating Cyber-Physical System-of-Systems (CPSoS) to function as a single unified system is the autonomy of its Cyber-Physical Systems (CPSs), which may lead to lack of coordination among CPSs and results in various kinds of conflicts. We advocate that to efficiently integrate CPSs within the CPSoS, we may need to adjust the autonomy of some CPSs in a way that allows them to coordinate their activities to avoid any potential conflict among one another. To achieve that, we need to incorporate the notion of governance within the design of CPSoS, which defines rules that can be used for clearly specifying who and how can adjust the autonomy of a CPS.  In this paper, we try to tackle this problem by proposing a new conceptual model that can be used for performing a governance-based analysis of autonomy for CPSs within CPSoS. We illustrate the utility of the model with an example from the automotive domain.

\end{abstract}


\section{Introduction}

 A Cyber-Physical System-of-Systems (CPSoS) can be defined as a System-of-Systems (SoS) that is composed of several independent and operable Cyber-Physical Systems (CPSs), which are networked together to achieve a certain higher goal.  Usually, a CPS is a system consisting of cyber components (e.g., the Electronic and Electric (E/E) systems of a vehicle), controlled components (e.g., physical objects such as other vehicles) and possibly of interacting humans (e.g., drivers, passengers) \cite{lollini2016amadeos}.

Assuring that a CPSoS/SoS can function as an integrated/unified system to support a common cause is a main objective for the CPSoS/SoS community \cite{Jamshidi2008,lollini2016amadeos}. However, such integration is not an easy task because of the special nature that distinguishes CPSoS/SoS from other types of systems, and especially the autonomy of its components (e.g., CPSs) \cite{lollini2016amadeos}. More specifically, besides ``directed'' CPSoS/SoS where CPSs have almost no autonomy, in ``acknowledged'', ``collaborative'' and ``virtual'' CPSoS/SoS  the autonomy of CPSs may lead to conflicts and unsafe situations due to the lack of coordination among one another. For instance, an autonomous vehicle hits and killed a woman that was walking outside of the crosswalk recently \cite{Levin2018}. This is an example where the autonomy of CPSs (e.g., an autonomous vehicle and a human) led to a lack of coordination among them that, in turn, has led to an unfortunate incident.

In a previous work \cite{gharibLADC2018}, we advocated that coordination among CPSs can be achieved by adjusting the autonomy level of some CPSs in a way that allows them to perform their activities without endangering other CPSs that are operating in the same environment. Although several researchers have suggested adjusting the autonomy level of a system based on various criteria such as its capability, motivations, behavior, etc. \cite{castelfranchi1995guarantees}. We proposed to determine the autonomy level of a CPS based on its \textit{Awareness} concerning its operational environment as well as its capability to safely perform its activity (e.g., \textit{Controllability}) \cite{gharibLADC2018}. Based on these criteria, a \textit{CPS} can have \textit{full, partial} or  \textit{limited autonomy} for performing a specific \textit{activity}.

However, we did not provide \textit{governance rules/policies} that specify who and how can adjust the autonomy of CPSs.  In other words, component systems (e.g., CPSs) maintain an ability to operate autonomously, but their operational mode is subordinated to a central managed purpose \cite{VanemanSoS2013,Morris2006}. Such central managed purpose can be expressed by \textit{governance rules/policies}. Governance can be defined as the set of rules, policies, and decision-making criteria that will guide the CPSoS/SoS while achieving its goals \cite{VanemanSoS2013}.  Governance is not a new concept, it is an emerging paradigm in Systems Theory \cite{KeatingSoS2014}, and it represents a cornerstone of an effective CPSoS/SoS \cite{VanemanSoS2013}. Despite this, it did not receive enough attention from the CPSoS/SoS community \cite{VanemanSoS2013,KeatingSoS2014}.  

In this paper, we argue that in order to efficiently integrate CPSs within their CPSoS, we need to incorporate the notion of governance within the CPSoS design.  In particular, we try to tackle this problem by proposing a new conceptual model that can be used for providing a governance-based autonomy analysis for CPSs within CPSoS. In other words, the model can be used for analyzing the autonomy level of CPSs taking into consideration governance rules defined by the CPSoS.



\begin{figure*}[!t]
\centering
\includegraphics[width=  0.94\linewidth]{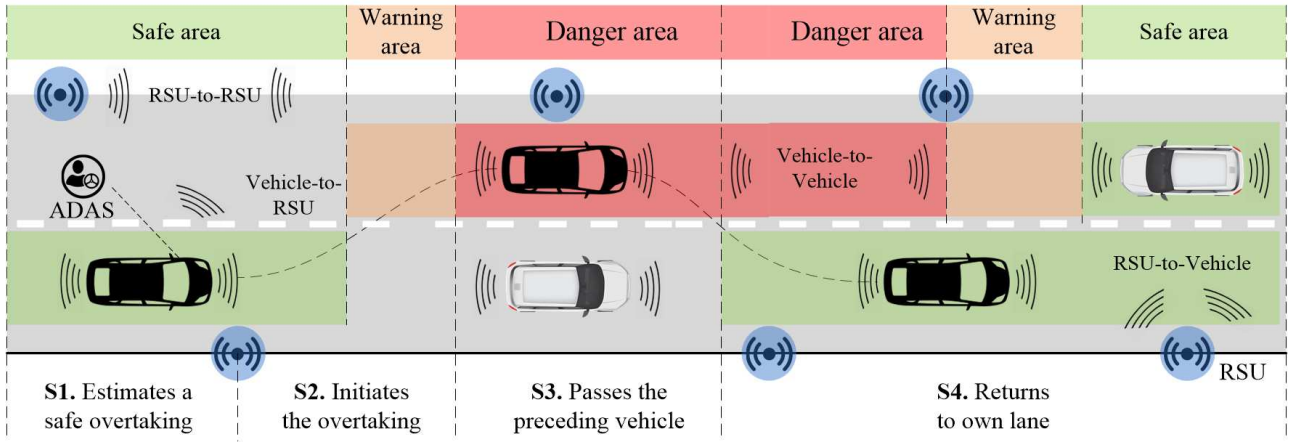}
\caption{A diagram of the cooperative overtaking assistance system with the critical zones}
\label{fig:overtakingSys}
\end{figure*}

 The remainder of this paper is organized as follows; Section II describes a motivating example we use to illustrate our work. We propose a conceptual model that can be used for providing a governance-based analysis of autonomy for CPSoS in Section III, and we illustrate its applicability to a realistic scenario from the automotive domain in Section IV. Related work is presented in Section V. Finally, we conclude and discuss future work in Section VI.

\section{Motivating Example}

 Overtaking is a complex and critical driving task, where a driver may make several decisions based on the traffic conditions \cite{FigueiraOver2020}, i.e., a driver needs to identify an acceptable size gap in the opposing traffic, the time at which he initiates the overtake as well as the time at which to return to its original lane in front of the preceding vehicle \cite{JenkinsOver2004}. In particular, overtaking is one of the major traffic safety problems, that is why there is much work towards developing driving support systems that reduce overtake-related accidents \cite{JenkinsOver2004,FigueiraOver2020}.


The cooperative driver overtaking assistance system aims at supporting drivers to avoid overtake-related accidents on undivided roads, where Advanced Driver Assistance Systems (ADAS), Road Side Units (RSUs), vehicles, and other road infrastructure cooperate to reduce overtake-related accidents. In particular, \textit{RSUs} collect and distribute information that assists drivers/ADAS to perform a safe overtaking. \textit{ADAS} aims at improving the driver's safety by a thorough task analysis of overtaking activity considering the driver's ability to complete a safe overtake. The ADAS can monitor, warn and even take control of the vehicle in case the driver is not able to perform/complete a safe overtake.  

Information can be exchanged between the system components either directly relying on wired or wireless channels, or indirectly relying on acquiring such information by sensing the domain. For example, RSUs/drivers can acquire information about close-by vehicles by sensing/seeing such vehicles \cite{gharibsose17}.

 The main components of the system are shown in Fig. \ref{fig:overtakingSys}, and we can also identify the four \textbf{S}teps in a successful overtake following \cite{Birk2009}: \textbf{S1.} the driver estimates the possibility of safely overtaking a preceding vehicle, \textbf{S2.} the driver initiates the overtaking, \textbf{S3.} the driver passes the preceding vehicle in the opposite lane, and \textbf{S4.} changing the lane back into the original lane of the vehicle, which completes the overtaking successfully. Considering these steps, a vehicle can be in i) a safe area, where there is no overtaking-related hazard; ii) warning area, where there is a hazard due to an overtake in progress; and iii) danger area, where there is an imminent danger due to an overtake in progress.

\section{A Conceptual model for Governance-based Analysis of Autonomy in Cyber-Physical Systems-of-Systems}
 
\begin{figure*}[!t]
\centering
\includegraphics[width=  0.72\linewidth]{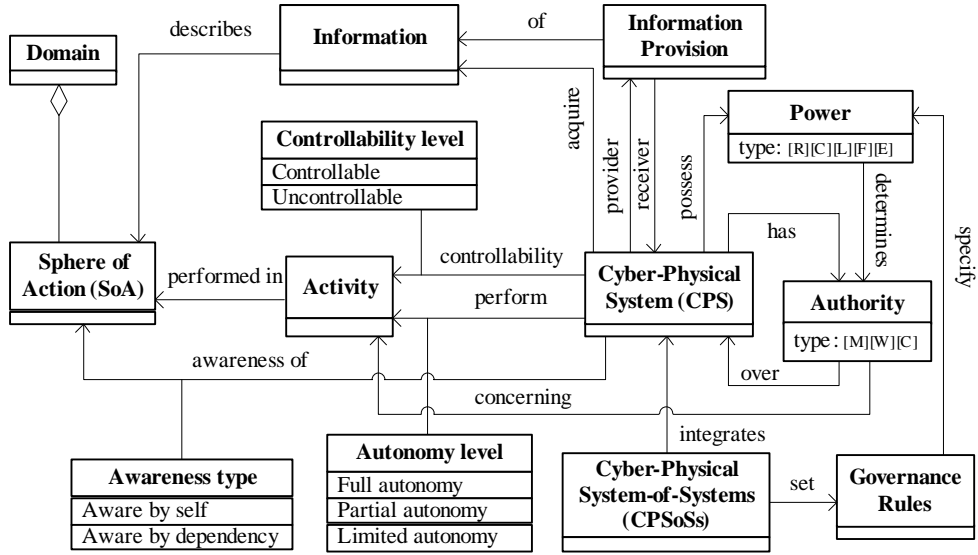}
\caption{The meta-model of the proposed conceptual model}
\label{fig:meta}
\end{figure*}

In this section, we present the conceptual model we developed for performing a governance-based analysis of autonomy levels for CPSs within CPSoS. This model builds on our previous work \cite{gharibLADC2018}, and extends it with concepts for modeling governance within CPSoS. The meta-model of our conceptual model is shown in Figure \ref{fig:meta}, where we can identify a \textit{CPSoS} that \textit{integrates} \textit{CPSs}. For instance, the cooperative driver overtaking assistance system is a \textit{CPSoS} that \textit{integrates} several  \textit{CPSs} such as RSUs, ADAS, drivers, etc. A \textit{CPS} can \textit{perform} \textit{activities} for achieving its own objectives and/or the objectives of the overall \textit{CPSoS}.  For example, a driver may \textit{perform} an overtake (an \textit{activity}) to pass a slower vehicle. Usually, an \textit{activity} is \textit{performed} in an operational environment (we call \textit{Sphere of Action (SoA)} \cite{gharibsose17}), which is a part of the domain/environment. For instance, an overtake (an \textit{activity}) can be \textit{performed} in part of the road (an \textit{SoA}).

A \textit{SoA} can be \textit{described} by \textit{information}. For example, an RSU can acquire information describing the situation of the traffic concerning some part of the road. \textit{CPSs} can rely on one another for information, i.e., a \textit{CPS} can provide/receive \textit{information} depending on the \textit{information provision} concept. For instance, a driver can depend on a RSU to provide him with information concerning the road situation.

A \textit{CPS} must be aware of its \textit{SoA} to operate in it, and the \textit{awareness of} relationship between a \textit{CPS} and a \textit{SoA} is used to capture such relation.  Following \cite{castelfranchi1995guarantees}, we differentiate between 1- \textit{aware by self}, a CPS has the self-capability to be aware of its \textit{SoA}, e.g., CPS is independent, and 2- \textit{ aware by dependency}, a CPS needs to depend on other CPS to be aware of its \textit{SoA}, e.g., CPS is dependent. For example, if a driver can acquire information describing the road situation by himself, he is \textit{aware by self}. While if he depends on an RSU for such information, he is \textit{aware by dependency.} 

 The \textit{controllability} of a \textit{CPS} over the performance of an \textit{activity} it aims to perform is captured relying on  the \textit{controllability}  relationship, which is characterized by one attribute,  namely \textit{controllability level} that can be: 1- \textit{Controllable}, a \textit{CPS} is able to detect and avoid any obstacle that might prevent it from safely performing its activity in a timely manner; 2- \textit{Uncontrollable}, a \textit{CPS} is not able to detect and/or avoid all obstacles that might prevent it from safely performing its activity in a timely manner.  For example, performing an overtake in clear visual conditions is \textit{controllable} by the driver since he has sufficient visibility for maintaining safe separation from other vehicles/obstacles while performing the overtake. While performing an overtake in unclear visual conditions and with no support of RSUs might be \textit{uncontrollable} by the driver. 


To capture the autonomy level of a \textit{CPS} concerning an activity performance, we extend the \textit{perform} relationship between the \textit{CPS} and \textit{Activity} concepts with the \textit{autonomy level} attribute that can be, 1- \textit{Full autonomy,} if the \textit{CPS}  is \textit{aware by self} of the environment, and the \textit{activity} is \textit{controllable} with respect to the \textit{CPS}  capability, 2- \textit{Partial autonomy,} if it is \textit{aware by dependency} and the \textit{activity} is \textit{controllable} by it, and 3- \textit{Limited autonomy,} if the \textit{activity} is \textit{uncontrollable} regardless if it is \textit{aware by self/dependency} of the environment.
 
In what follows, we describe the concepts, relationships and attributes that can be used for modeling governance within CPSoS. A \textit{CPSoS} can set \textit{Governance rules}, which can be defined as a set of rules, policies, and decision-making criteria that will guide the CPSs while achieving their goals \cite{VanemanSoS2013}. For example, adjusting the autonomy of the driver (a \textit{CPS}) from \textit{Full autonomy} to  \textit{Partial} or \textit{Limited autonomy} based on his type of \textit{awareness} of the \textit{SoA} and his \textit{controllability level} concerning the activity can be specified within the \textit{Governance rules} specified by the Cooperative Driver Overtaking Assistance System (a \textit{CPSoS}).

\textit{Governance rules} can \textit{specify} the \textit{power} a \textit{CPS} may \textit{possess}, where  \textit{power} can be defined as the capacity or ability to direct or influence the behavior of others \cite{french1959bases,dictionary1989oxford}, i. e., the power of a \textit{CPS} within the \textit{CPSoS} is the maximum potential ability of a \textit{CPS} to influence the behavior of other \textit{CPS} concerning some performed activity. Following \cite{french1959bases}, we adopt five bases/sources of power: 1- \textbf{R}eward power is defined as power whose basis is the ability to reward; 2- \textbf{C}oercive power is defined as power whose basis is the ability to punish; 3- \textbf{L}egitimate power is defined as power whose basis is a formal authority that an individual has, which allows it to influence another individual(s), who has/have an obligation to accept such influence; 4- Re\textbf{F}erent power is defined as power whose basis is trust, respect, and admiration between individuals; and 5- \textbf{E}xpert power is defined as power whose basis is knowledge and experience that an individual attributes to another one within a specific area.

 \textit{Power} \textit{determines} the \textit{authority} a \textit{CPS} may \textit{has} over another \textit{CPS} concerning the performance of some activities, where \textit{authority} can be defined as the right to give orders, make decisions, and enforce obedience \cite{dictionary1989oxford}. For instance, the ADAS (a \textit{CPS}) can \textit{possess} an \textbf{E}xpert/\textbf{L}egitimate power  over the driver (another \textit{CPS}). This power gives the ADAS the \textit{authority} over the driver performance \textit{concerning} the overtake activity.

We differentiate between three types of \textit{authorities\footnote{These types are not mutually exclusive, i.e., a \textit{CPS} may have all three types of \textit{authorities} over another \textit{CPS}}}: 1- \textbf{M}onitoring is the process of observing and analyzing the behavior of an individual in order to detect any undesirable behavior; 2- \textbf{W}arning is the process of informing an individual, usually in advance, of possible danger, problem, or other unpleasant situation; and 3- \textbf{C}ontroling is the process of influencing, directing or even determining the behavior of an individual.  Several researchers (e.g., \cite{french1959bases,rodin1979social,yukl1991importance}) have concluded that various sources of power have different influence over the individuals' behavior, which is out of the scope of this paper. In this work, we consider \textbf{E}xpert and \textbf{L}egitimate power, where the first grants only \textbf{M}onitoring and \textbf{W}arning \textit{authorities}, while the last grants \textbf{M}onitoring, \textbf{W}arning as well as \textbf{C}ontroling \textit{authorities}.

\begin{figure*}[!t]
\centering
\includegraphics[width=  0.99\linewidth]{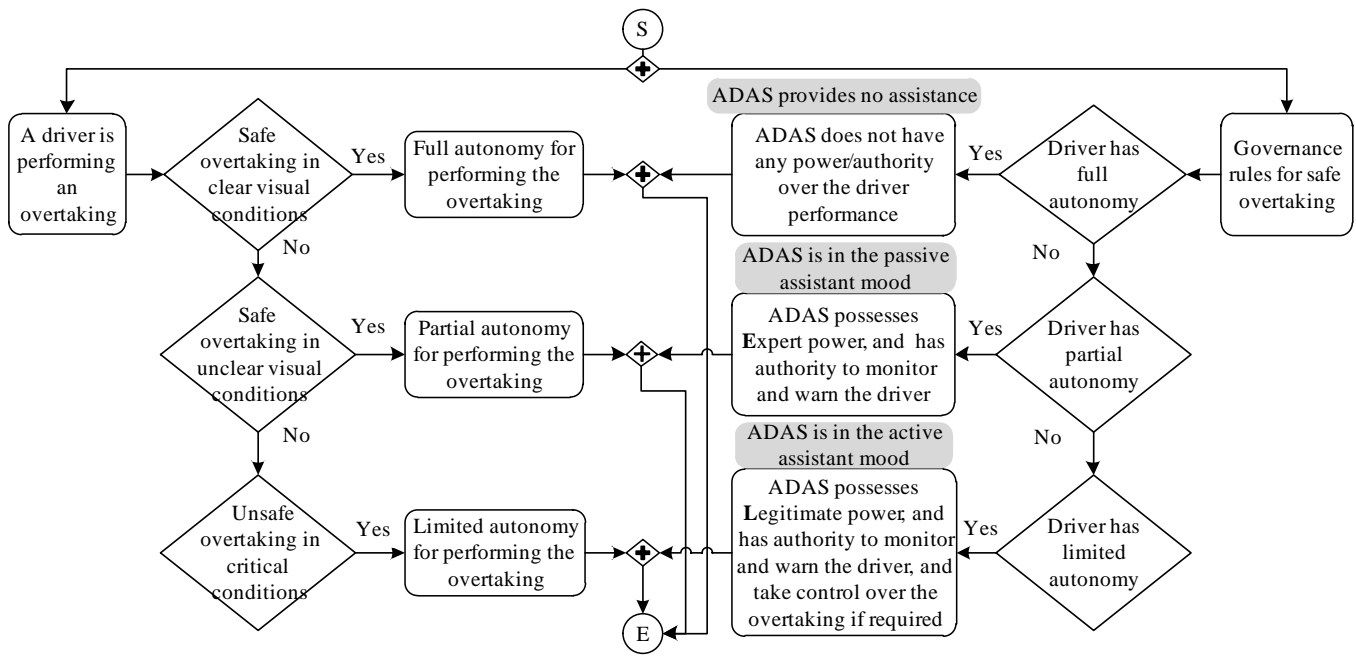}
\caption{An abstract flow chart of a governance-based analysis of the driver's autonomy}
\label{fig:flowchart}
\end{figure*}

For example, when the ADAS have the \textbf{E}xpert power, it can be in the passive overtaking assistant mode, i.e., it has the monitoring and warning \textit{authorities} over the driver when the driver has a partial autonomy to perform an overtake. While when the ADAS have \textbf{L}egitimate power, it can be in the active overtaking assistant mode, i.e., it has the monitoring, warning and also controlling\footnote{Several ADAS has been designed to increase the driver's safety by interrupting and controlling the activity to be performed (overtaking) \cite{gharibcritis17}.} \textit{authorities} over the driver when the driver has a limited autonomy to perform an overtake.


\section{Illustrating  the utility of the conceptual model}

We illustrate the utility of the conceptual model by applying it to a realistic scenario concerning the Cooperative Driver Overtaking Assistance System. Consider for example a driver that aims at reaching his/her destination safely using an undivided (two-lanes) road.  Depending on the situation of the traffic, the driver may perform several overtakes before reaching his destination. Most overtakes on undivided roads can be broadly classified under three different types:

\begin{enumerate}

\item \textit{Safe overtaking in clear visual conditions,} the driver has sufficient visibility for maintaining safe separation from other vehicles/obstacles while performing the overtake, i.e., the driver can self-detect  (\textit{aware by self}) and avoid any vehicle/obstacle while performing the overtake (the overtake is \textit{controllable}). 

\item \textit{Safe overtaking in unclear visual conditions,} the driver may not has the self-capability to detect other vehicles while performing his overtake. This could be due to the nature of the road (e.g., a sharp curve) or due to weather conditions (e.g., fog, heavy rain, etc.). In such a situation, the driver can rely on RSUs to provide him with such information (i.e., \textit{aware by dependency}). However, with such information, the overtake is considered \textit{controllable} by the driver.  

\item \textit{Unsafe overtaking in critical conditions}, the driver is considered incapable of performing a safe overtake regardless of his type of awareness of the \textit{SoA} (e.g., \textit{aware by self} or \textit{aware by dependency}). Note that the cooperative driver overtaking assistance system identifies such overtakes by analyzing the location, speed and direction of other vehicles in the maneuver area, i.e., the system can estimate whether an overtake is \textit{controllable} or \textit{uncontrollable} by the driver.

\end{enumerate}

Taking the previous three types of overtaking, the cooperative driver overtaking assistance system and to increase drivers' safety by reducing overtake-related accidents can set \textit{Governance rules} for specifying the autonomy allowed to drivers based on their type of \textit{awareness} of the \textit{SoA} and their \textit{controllability levels} concerning the overtakes. Such rules can be interpreted into power that determines authorities over the driver's performance concerning the overtake activity.

For the first type of overtaking, the driver can have \textit{Full autonomy} concerning any overtake he wishes to perform, i.e., the ADAS system is not granted any power/authority over the driver and it provides no assistance at all. In the second type of overtaking, the driver can have \textit{Patial autonomy} concerning any overtake he wishes to perform. The ADAS system is granted an \textit{Expert power} over the driver, which allows it to be in the passive assistance mode, i.e., it has the authority to monitor the driver's behavior and warn him about any possible dangerous situation. While in the last type of overtaking, the driver can have \textit{Limited autonomy} concerning any overtake he wishes to perform. The ADAS system is granted a \textit{Legitimate power} over the driver, which allows it to be in the active assistance mode, i.e., it has the authority not only to monitor and warn the driver but also to interrupt and control the overtake (e.g., reduces speed, applies breaks, prevents initiating the overtake, prevents changing the lane). Fig. \ref{fig:flowchart} shows an abstract flow chart of a governance-based analysis of the driver's autonomy concerning the three different types of overtaking.

\begin{figure*}[!t]
\centering
\includegraphics[width=  0.68\linewidth]{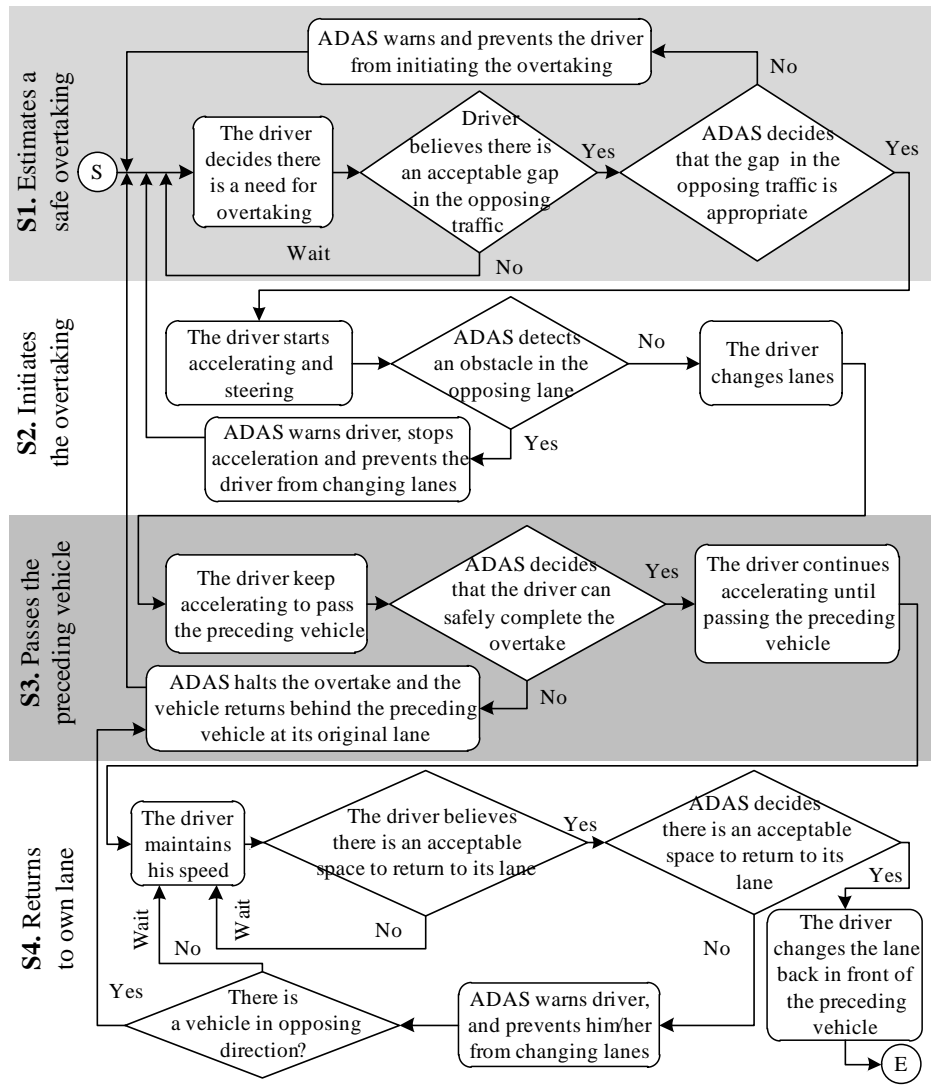}
\caption{An flow chart of a governance-based analysis of the driver's autonomy in \textit{Unsafe overtaking in critical conditions}}
\label{fig:taskunsafe}
\end{figure*}

Due to space limitation, we only describe the task analysis concerning the driver's autonomy in unsafe overtaking in critical conditions that is shown in Fig. \ref{fig:taskunsafe}. As previously mentioned, a successful overtake in undivided roads consists of four main \textbf{S}teps.

In \textbf{S1}, the driver first decides there is a need for overtaking, which depends on the speed of the preceding vehicle, his/her desired speed, etc. After deciding there is a need for overtaking, the driver waits an acceptable gap in the opposing traffic that allows him/her to initiate the overtake. However, even if the driver believes that there is an acceptable gap in the opposing traffic, the ADAS may have another opinion as drivers might have poor judgment concerning the distance and speed of opposing vehicles, or they might not even see such vehicles until they initiate the overtake. In other words, the ADAS have better judgment than the driver, therefore, if the ADAS decides that the gap is not appropriate, it will warn the driver about that. If the driver did not comply with the warning, the ADAS will prevent him/her from initiating such overtake. In particular, the driver is allowed to initiate the overtake only if the ADAS allows that. 

In \textbf{S2}, the driver starts accelerating and steering aiming at changing lanes, if the ADAS detects any obstacle/vehicle in the opposing lane that may prevent the driver from safely changing lanes, it warns the driver, stops acceleration and prevents the driver from changing lanes if it has to, i.e., the ADAS halts the overtake and the driver stays behind the preceding vehicle. Otherwise, the driver is allowed to change lanes. The driver starts \textbf{S3} by accelerating to pass the preceding vehicle, yet if the ADAS decided that the driver cannot safely complete the overtake, it halts the overtake and the vehicle returns behind the preceding vehicle.

After passing the preceding vehicle, \textbf{S3} completes and  \textbf{S4} starts. In which, the driver maintains his/her speed trying to find an acceptable space to return to its original lane in front of the preceding vehicle(s). However, the driver is allowed to do that only if the ADAS decides that the available gap is adequate for completing the overtake. Otherwise, the ADAS warns the driver that the space is not sufficient/adequate, and if the driver tries to change lanes, the ADAS will prevent that. In such case, the driver can maintain his/her speed and stay at the opposing lane waiting for adequate space/gap if there is no vehicle in the opposing direction. While if there is a vehicle, the ADAS will halt the overtake and the vehicle will return behind the preceding vehicle.

\section{Related work}



Several researchers have devoted effort toward researching governance for CPSoS/SoS. For instance, Morris et al. \cite{Morris2006} survey the available literature concerning information technology (IT) governance and identify six key characteristics of good IT governance that can be used for SoS. Moreover, Vaneman and Jaskot   \cite{VanemanSoS2013} worked toward developing a criteria-based framework for SoS governance by conducting a survey of governance practices within the IT community with the main aim of identifying elements of good SoS governance. Keating \cite{KeatingSoS2014} explores the implications of Complex System Governance (CSG) trying to find suggestions or even solutions for similar governance challenges faced in the development of the System of Systems Engineering (SoSE) area. Based on \cite{KeatingSoS2014}, Keating and Bradley \cite{keating2015complex} presented a preliminary reference model suitable for the emerging field of CSG.

Unlike existing solutions, we propose to link the governance concept to the concepts of power and authority when adjusting the level of autonomy of some CPS within the CPSoS.

\section{Conclusions and Future Work}

In this paper, we advocate that to efficiently integrate CPSs within the overall context of their CPSoS, we need to incorporate the notion of governance within the CPSoS design, which defines rules that specify who and how can adjust the autonomy of a CPS.  Moreover, we proposed a conceptual model that can be used for performing a governance-based analysis of autonomy for CPSs within CPSoS. Additionally, we illustrated the utility and applicability of our model by applying it to a realistic example from the automotive domain.


For future work, we intend to investigate the properness of the proposed controllability levels, i.e., whether we need to consider more than only two levels. Moreover, we plan to extend the autonomy levels we considered to cover all 6 levels of driving automation ranging from 0 (fully manual) to 5 (fully autonomous) defined by the Society of Automotive Engineers (SAE).  We will also better validate our model by applying it to other systems from different domains (e.g., rail transport systems, smart grid, Industry 4.0, etc.). 

\section*{Acknowledgment}

This work has received funding the European Union's Horizon 2020 research and innovation program under the Marie Sklodowska-Curie grant agreement No. 823788 - ADVANCE.




\end{document}